# WGMR Self-Injection Locking Method Based on Enhanced Optical Feedback with Auxiliary Prism

Wu Jiajun[1,2], Zhong Shan[1*], Kang Songbai[1*]

[1] *Key Laboratory of Atomic Frequency Standards,*

*Innovation Academy for Precision Measurement Science and Technology,*

*Chinese Academy of Sciences, Wuhan 430071, China*;

[2] *University of Chinese Academy of Sciences, Beijing 100049, China*

\* *kangsongbai@apm.ac.cn* & *zhongshan@apm.ac.cn*

**Abstract**

The optical feedback intensity is an important parameter for realizing narrow linewidth lasers in Whispering-gallery-mode resonator (WGMR) self-injection locking technology. We proposed an approach that enhances the intensity of intracavity feedback in crystalline WGMR by using only a single coated auxiliary prism. Compared to the Rayleigh scattering, the feedback intensity of the enhanced scheme increased by more than a hundred times. Furthermore, we demonstrated that, with the enhanced approach, the instantaneous linewidth of the laser was suppressed to 7 Hz level, the locking range was expanded up to 8 GHz, and the relative intensity noise (RIN) was reduced to -152 dBc/Hz@10MHz. The feedback enhanced design is compact, easy-to-operated and can be integrated with the WGMR. It provides a miniaturized solution for controlling optical feedback intensity in WGMR self-injection locking technology.

**Objective**

Narrow linewidth lasers based on WGMR self-injection locking (**SIL**) [1,2] have the potential to be used in fields such as coherent optical communication, optical atomic clocks, high-resolution spectroscopy, and precision frequency measurement. The intensity of SIL optical feedback is a crucial parameter that determines the performance of the locked laser, including instantaneous linewidth, locking range, and relative intensity noise (**RIN**). At present, the SIL technique is commonly based on WGMR's Rayleigh backscattering (**RB**) to create optical feedback [3]. However, the optical feedback intensity is uncontrollable. There have been fruitful investigations about the feedback control for the on-chip WGMRs such as $Si_3N_4$ and AlN by using micro-cavity refectors [4,5]. However, there are few studies on how to control the feedback intensity for the high Q fluoride crystalline WGMRs. Liang Wei et al. proposed a feedback intensity enhancement scheme for crystalline WGMRs by using a set of optical components including a coupling prism, collimation lenses, and a mirror reflector [6]. However, the additional components increase the complexity and instability of the system. In this work, we propose a compact and easy-to-operate feedback enhancement approach for the crystalline WGMR by replacing the set of feedback-enhanced components with a designed auxiliary prism, which can be integrated with the WGMR. It provides a compact solution for controlling feedback intensity in WGMR SIL technology.

**Methods**

We designed an auxiliary prism to enhance the counterclockwise (CCW) light amplitude inside the WGMR(**Figure 1(a)**). The auxiliary prism first extracts the clockwise (CW) light out from the WGMR with a specific angle Φ that is determined by the coupling phase-matching condition. The coupled output beam travels in the prism with an optical length of d, then is vertically reflected by high reflective film coated on the prism, and finally coupled back into the WGMR along the original trace to enhance the CCW light amplitude with any additional optical components. The bottom angle of the auxiliary prism is crucial and is designed to equal angle Φ for high back-coupling efficiency. In the experiment, we used a homemade high Q ($2\times10^9$) MgF$_2$ WGMR (**Figure 1(b)**) as the platform and an H-ZF13 coated auxiliary prism to enhance the TE mode's CCW light amplitude. The prism's bottom angle is designed to be 52.3 (for TE mode@1.55 μm). The output coupling point is also important and should ensure that d is approximately equal to the light's Rayleigh length to prevent beam divergence. So, the WGMR and auxiliary prism's coupling point is chosen to be about 500 μm from the bottom corner of the prism.

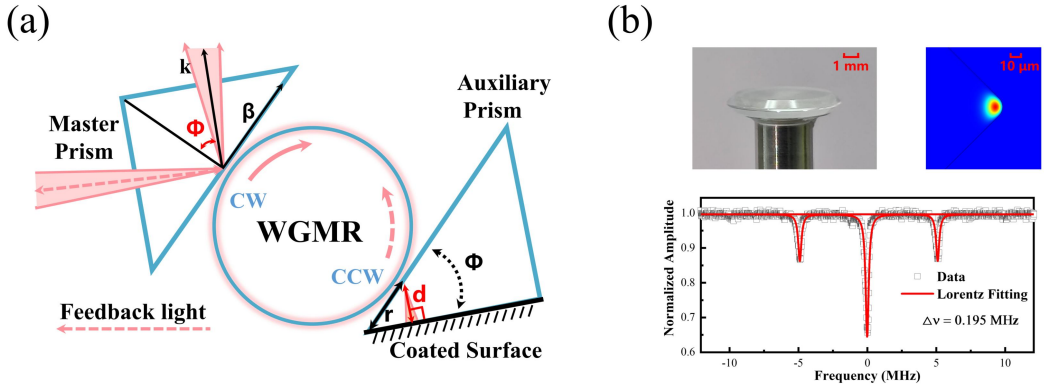

Fig. 1 (a) Schematic diagram of enhancing WGMR optical feedback intensity through auxiliary prism; (b) Magnesium fluoride WGMR and its fundamental mode size simulation and Q factor measurement.

To quantify the CCW light enhancement, we used an optical circulator to monitor the feedback light intensity and compare the intensities with and without the auxiliary prism enhancement (**Figure 2(a)**). Furthermore, we set up a WGMR SIL laser to verify the performance improvement with the feedback enhancement approach (**Figure 2(b)**).

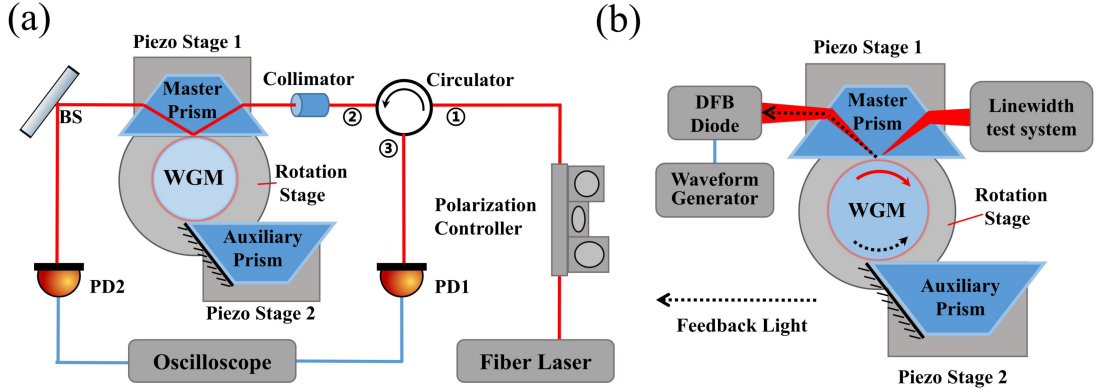

Fig. 2 (a) Diagram for measuring feedback enhancement by auxiliary prism; (b) Diagram of self-injection locking laser with auxiliary prism.

**Results and Discussions**

Compared to the WRGR's Rayleigh backscattering, the auxiliary prism improved the WGMR's CCW intensity for the TE mode by two orders of magnitude (**Figure 3(a)**), and did not enhance the TM mode feedback intensity(**Figure 3(b)**). If the prism's bottom angle can continue to be optimized, the feedback intensity can be further increased.

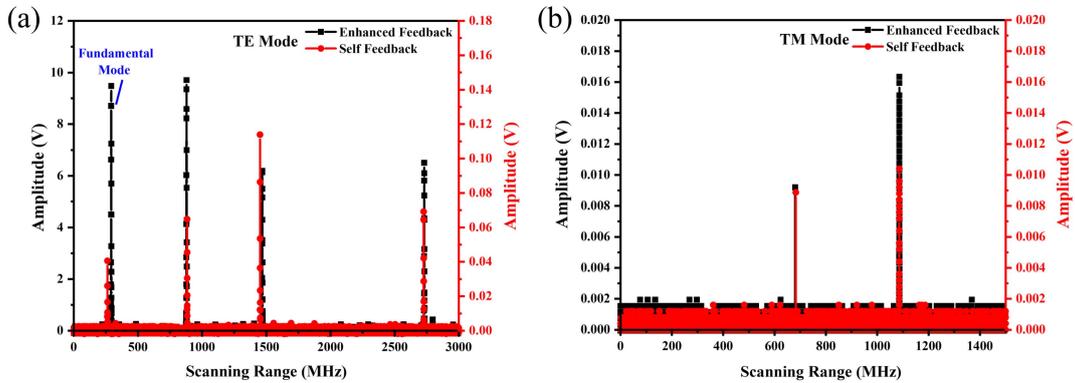

Fig. 3 The auxiliary prism enhances the feedback intensity for different modes of the WGMR. (a) TE mode; (b) TM mode

After the enhanced feedback approach was applied to a WGMR SI locking laser, the results show that the locked laser's performance was improved. For the instantaneous frequency noise (**Figure 4(a)**), the free-running case was about $3.5 \times 10^5$ $Hz^2/Hz$ ($\approx 1.1\ MHz$), the RB SI locking case was about 40 $Hz^2/Hz$ ($\approx 125\ Hz$). And the enhanced feedback SIL case was 2.5 $Hz^2/Hz$ ($\approx 7\ Hz$), obtaining almost 50 dB and 10 dB linewidth suppression gain, respectively. For the SIL range (**Figure 4(b)**), the RB SIL case was about 0.8 GHz and the enhanced feedback SIL case expanded the range up to 8 GHz, substantially enhancing the injection locking robustness. For the RIN (**Figure 4(c)**), the free-running case was about -142 dBc/Hz at 10 MHz, the RB SI locking case was about -147 dBc/Hz at 10 MHz, and the enhanced feedback SI locking suppressed the RIN to -152 dBc/Hz at 10 MHz.

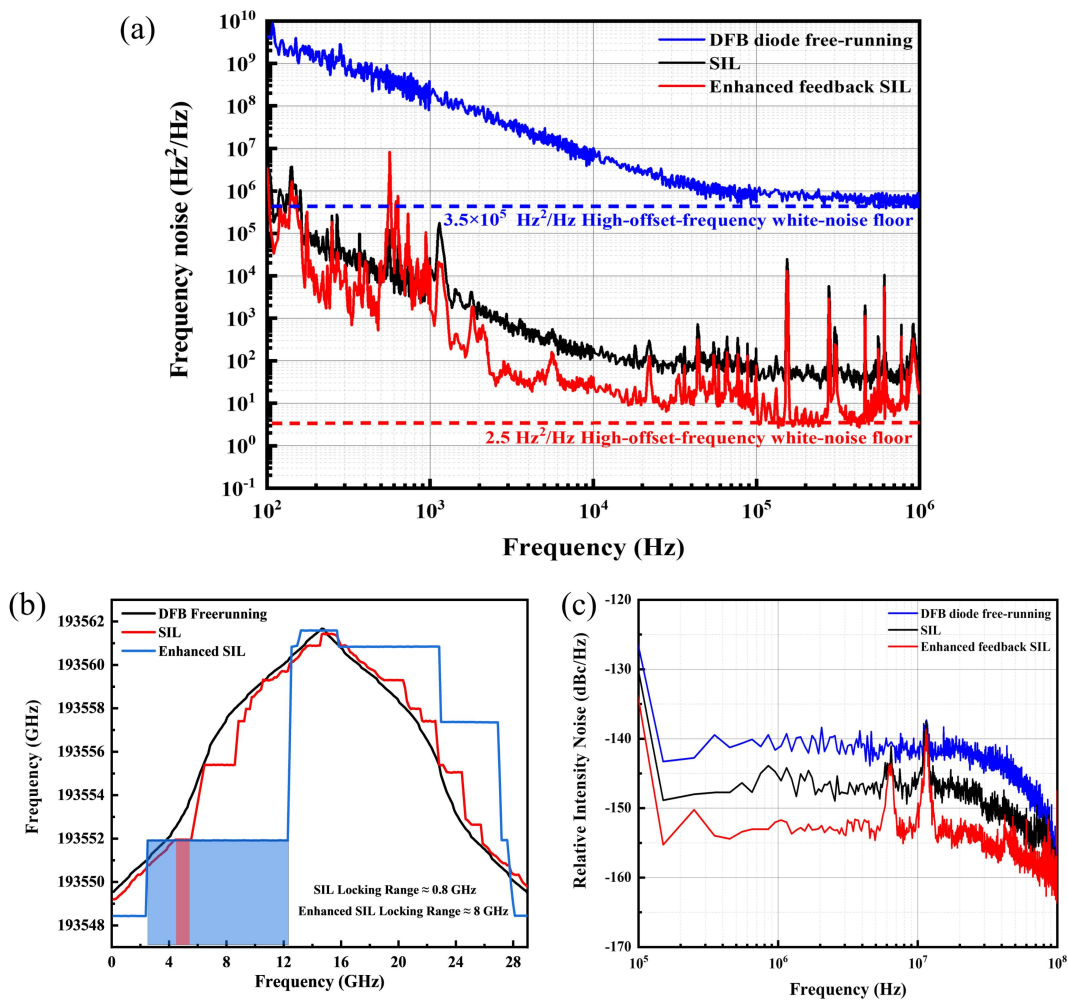

Fig. 4 (a) Frequency noise of DFB laser free running (blue line), cavity Rayleigh scattering self-injection locking (black), auxiliary prism enhanced feedback self-injection locking (red); (b) the laser locking range based on Rayleigh scattering (red) and auxiliary prism enhanced feedback self-injection locking (blue); (c) RIN performances of DFB free running (blue), Rayleigh scattering self-injection locking (black), and auxiliary prism enhanced feedback self-injection locking (red).

**Conclusions**

We proposed a crystalline WGMR feedback enhancement approach with an auxiliary prism and successfully demonstrated the CCW light intensity in the WGMR by more than two orders of magnitude. With the auxiliary prism enhancement, the WGMR SIL laser performance, including the instantaneous linewidth, the locking range, and RIN have been significantly improved, compared to the free-running and the RB SI locking cases. The proposed approach provides a compact solution for controlling feedback intensity in WGMR , especially suitable for long-wavelength SIL lasers where the resonator's RB amplitude is low.

## Acknowledgements

The authors gratefully acknowledge the suport from CAS and thank Dr. Chicheng Che for help on fabricating the high Q WGMR.